\documentclass[aps,prl,reprint,twocolumn,superscriptaddress,showpacs,showkeys,floatfix,bibnotes]{revtex4-2}

\usepackage{graphicx}
\usepackage{epstopdf}
\usepackage{amssymb}
\usepackage{amsmath}
\usepackage{bm}
\usepackage{braket}
\usepackage{color}

\usepackage[colorlinks,citecolor=blue]{hyperref}

\begin{document}
	
	\title{Dissipative Nonlinear Thouless Pumping of Temporal Solitons}
	
	\author{Xuzhen Cao}
	\affiliation{State Key Laboratory of Quantum Optics and Quantum Optics Devices, Institute of Laser Spectroscopy, Shanxi University, Taiyuan 030006, China}
	\affiliation{Collaborative Innovation Center of Extreme Optics, Shanxi University, Taiyuan 030006, China}

	\author{Chunyu Jia}
	\affiliation{Department of Physics, Zhejiang Normal University, Jinhua 321004, China}
	
	\author{Ying Hu}
	\thanks{huying@sxu.edu.cn}
	\affiliation{State Key Laboratory of Quantum Optics and Quantum Optics Devices, Institute of Laser Spectroscopy, Shanxi University, Taiyuan 030006, China}
	\affiliation{Collaborative Innovation Center of Extreme Optics, Shanxi University, Taiyuan 030006, China}

	\author{Zhaoxin Liang}
	\thanks{zhxliang@zjnu.edu.cn}
	\affiliation{Department of Physics, Zhejiang Normal University, Jinhua 321004, China}
	
\begin{abstract}The interplay between topology and soliton is a central topic in nonlinear topological physics. So far, most studies have been confined to conservative settings. Here, we explore Thouless pumping of dissipative temporal solitons in a nonconservative one-dimensional optical system with gain and spectral filtering, described by the paradigmatic complex Ginzburg-Landau equation. Two dissipatively induced nonlinear topological phase transitions are identified. First, when varying dissipative parameters across a threshold, the soliton transitions from being trapped in time to quantized drifting. This quantized temporal drift remains robust, even as the system evolves from a single-soliton state into multi-soliton state. Second, a dynamically emergent phase transition is found: the soliton is arrested until a critical point of its evolution, where a transition to topological drift occurs. Both phenomena uniquely arise from the dynamical interplay of dissipation, nonlinearity and topology.
\end{abstract}
	
	\maketitle
	
The fascinating synergy of topology and interaction enables new paradigms for exploring topological transport~\cite{Kraus2012,Wang2013,Lohse2016,Nakajima2016,Lu2016,Yongguan2016,Lohse2018,Zilberberg2018,Wenchao2018,Cerjan2020,Nakajima2021,Minguzzi2022,Bastidas2022} and its applications~\cite{Citro2023}. An emblematic instance concerns Thouless pumping~\cite{Thouless1982,Thouless1983,Niu1984,Di2010}, where slow periodic variations in system parameters induce quantized shift of particles that follows Chern number of the underlying Bloch bands. Although topological pump was originally introduced for essentially linear systems, recent experiments in optical waveguides demonstrated quantized pumping even for strong nonlinearities~\cite{Jurgensen2021}. Here, nonlinearity acts to quantize transport via the formation of solitons, which remain identical after each period (up to translation invariance), and symmetry-breaking bifurcations. Nowadays, nonlinear topological physics~\cite{Smirnova2020} is a burgeoning field with many theoretical ~\cite{Torio2001,Nakagawa2018,Mostaan2022,Qidong2022,Fu2022,Jurgensen2022,Bertok2022,Jurgensen2023,Jurgensen2023,Aligia2023,Tuloup2023,Cao2024,HaoarXiv2024} and experimental~\cite{Bongiovanni2021,Walter2023,Viebahn2024} developments, in platforms from nonlinear optics~\cite{Jurgensen2021}, photonics~\cite{Bongiovanni2021} to interacting quantum gases~\cite{Walter2023,Viebahn2024}. So far, many studies have been confined to conservative settings. Instead, below we investigate nonlinear Thouless pumping of dissipative solitons (DS)~\cite{AkhmedievBook} in nonconservative (non-Hermitian), nonlinear systems.

       \begin{figure}[t]
    	\centering
    	\includegraphics[width=1\columnwidth]{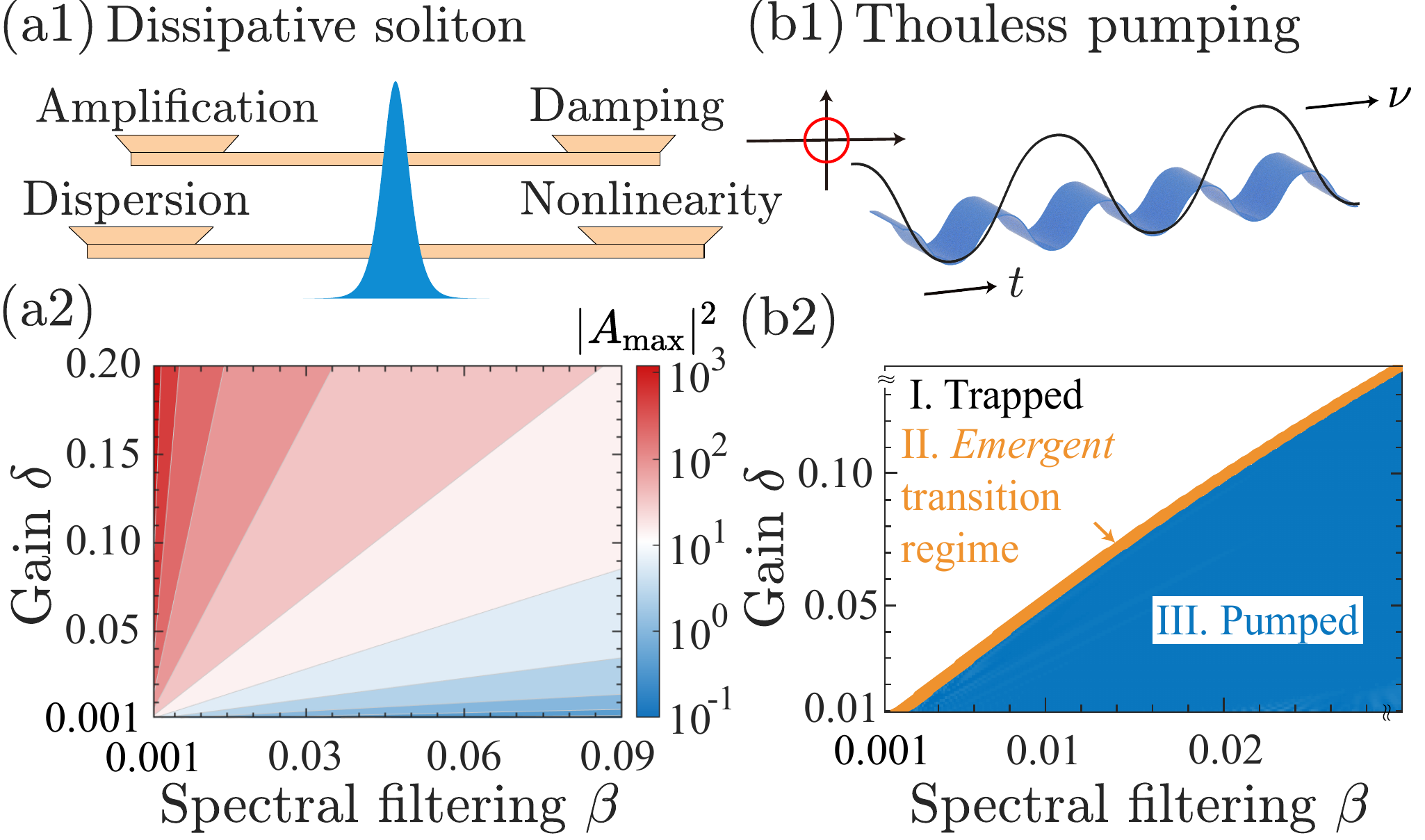}
    	\caption{Dissipative nonlinear Thouless pumping of temporal DS. (a1) Schematic of a DS sustained by double balances. (a2) Peak intensity of DS [Eq.~(\ref{DS})] as a function of spectral filtering $\beta$ and linear gain $\delta$, when the nonlinear gain is $\epsilon=0.8\epsilon_p$. (b1) Schematic of Thouless pump [Eq.~(\ref{OSL})]. (b2) Illustration of phase diagram of DS pumping. }\label{Fig1}
    \end{figure}
                            
      DSs are self-localized structures in complex nonconservative systems such as lasers~\cite{Renninger2008,Grelu2012,Peng2019,Helgason2021}, microresonantors~\cite{Lucas2017,Nikola2024} and microcavity polaritons~\cite{Sich2012}, and are building blocks in innovative laser designs and optical information processing~\cite{Kippenberg2018}. Unlike conservative soliton, DS is fundamentally dissipative, relying on the balances not only  between nonlinearity and dispersion but also between gain and loss [Fig.~\ref{Fig1}(a)]. Recently, topological phenomenology in open nonlinear systems have garnered numerous interests. In particular, topological lasing~\cite{Harari2018}, pumping of Bogoliubov quasiparticles~\cite{Ravets2024}, and topological classification of driven-dissipative nonlinear systems~\cite{Villaarxiv2024} have been reported. However, exploring the interplay between topology and DS remains elusive. 
      
      \begin{figure*}[t]
		\centering
		\includegraphics[width=0.9\textwidth]{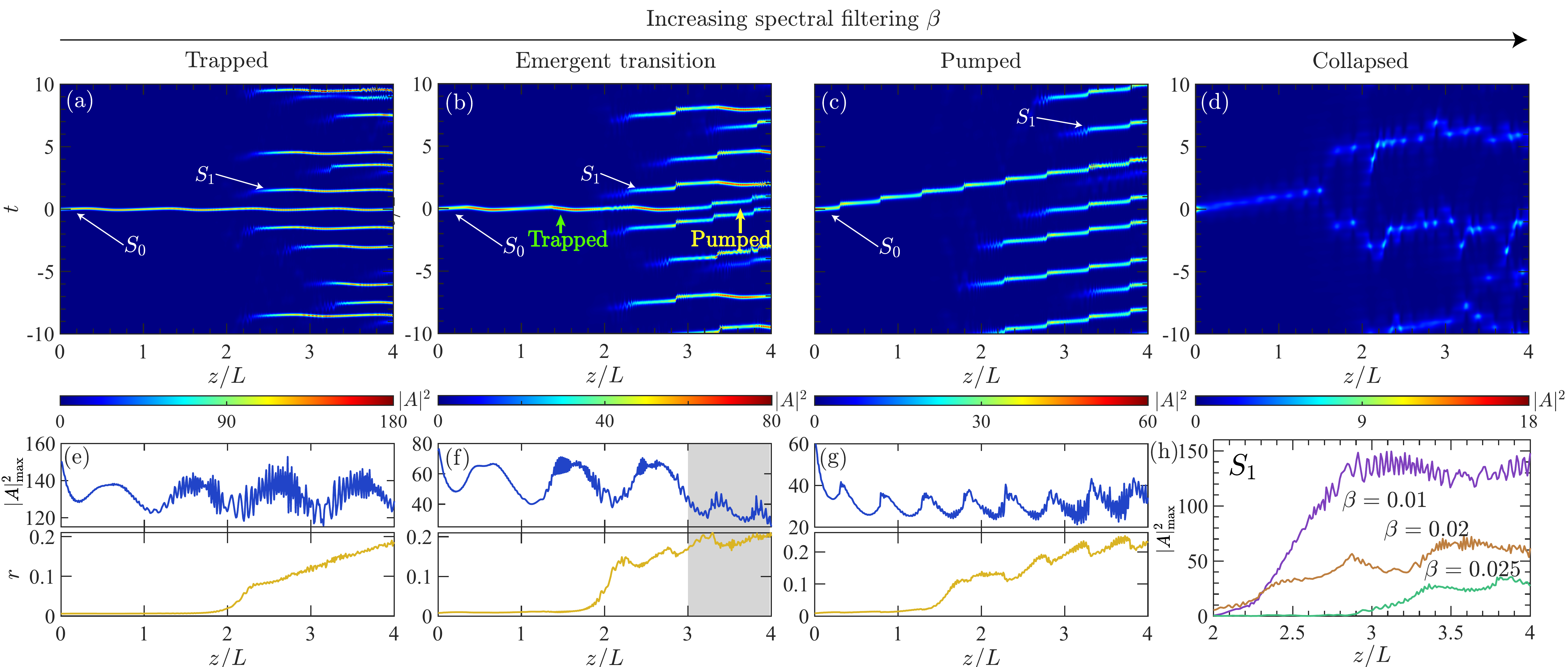}
		\caption{Four distinct regimes of DS pumping dynamics. The spectral filtering is (a) and (e) $\beta=0.01$, (b) and (f) $\beta=0.02$, (c) and (g) $\beta=0.025$, (d) and (h) $\beta=0.1$. Results are obtained by numerically~\cite{Supple} solving Eq.~(\ref{CGLE}) with an initial soliton given by Eq.~(\ref{DS}). We take linear gain $\delta=0.1$ and nonlinear gain $\epsilon=0.8\epsilon_p$. For Thouless pump (\ref{OSL}), we use $V_s=V_l=5$ and $\nu=0.1$. In (a)-(d), we use $dz=10^{-5}$, $dt=0.0025$ and a system size $N_T=41$ with periodic boundaries. (a)-(d) show intensity distributions $|A(t,z)|^2$. Panels (e)-(g) show evolutions of peak intensity $|A_\text{max}(t,z)|^2$ and participation ratio $r(t,z)$ corresponding to (a)-(c), respectively. (h) Evolutions of the peak intensity of satellite soliton $S_1$ in (a)-(c), respectively. }\label{Fig2}
	\end{figure*}

      Given the pivotal role that gain and loss play in DS formation, open challenges arise regarding whether DS pumping can be quantized from a dynamical point of view: The pump disturbs the energy balance between DS and its environment, and the ensuing dissipative dynamics in turn feeds back into the system via nonlinearity, so it is far from clear whether the DS can reproduce itself at the end of each pump cycle. Moreover, gain and loss in the medium entail non-Hermitian band structures, which potentially have distinct topological properties from their Hermitian counterparts~\cite{Kawabata2019,Bergholtz2021}.
    
Here, we consider a one-dimensional nonlinear optical system with gain and spectral filtering described by the paradigmatic complex Ginzburg-Landau equation (CGLE)~\cite{Aranson2002}. By numerically and analytically studying Thouless pumping [Fig.~\ref{Fig1}(b1)] of a temporal DS (i.e., localized structure in time), we identify two distinct topological phase transitions induced by dissipative mechanisms [Fig.~\ref{Fig1}(b2)]. (i) When modulating spectral filtering or gain beyond a threshold, DS can transition from being trapped to topologically quantized drifting in time domain. This quantized shift is remarkably robust, persisting even as the system evolves from a single-soliton state into a multi-soliton state. (ii) A dynamically emergent topological transport occurs for certain parameters: DS remains trapped in time until a critical point of its propagation, after which it exhibits a topological temporal shift. Unlike scenario (i), where the transition is driven by variations of system parameters, it is the dynamics itself that induces topological phase transition in (ii). Both phenomena uniquely arise from the \textit{dynamical} interplay of dissipation, nonlinearity and topology, which acts through the dual balances sustaining DSs and dynamical reshaping of soliton profiles. This work lays the groundwork for exploring dissipative nonlinear topology in a broad range of systems described by CGLE. 

\begin{figure*}[t]
		\centering
		\includegraphics[width=0.9\textwidth]{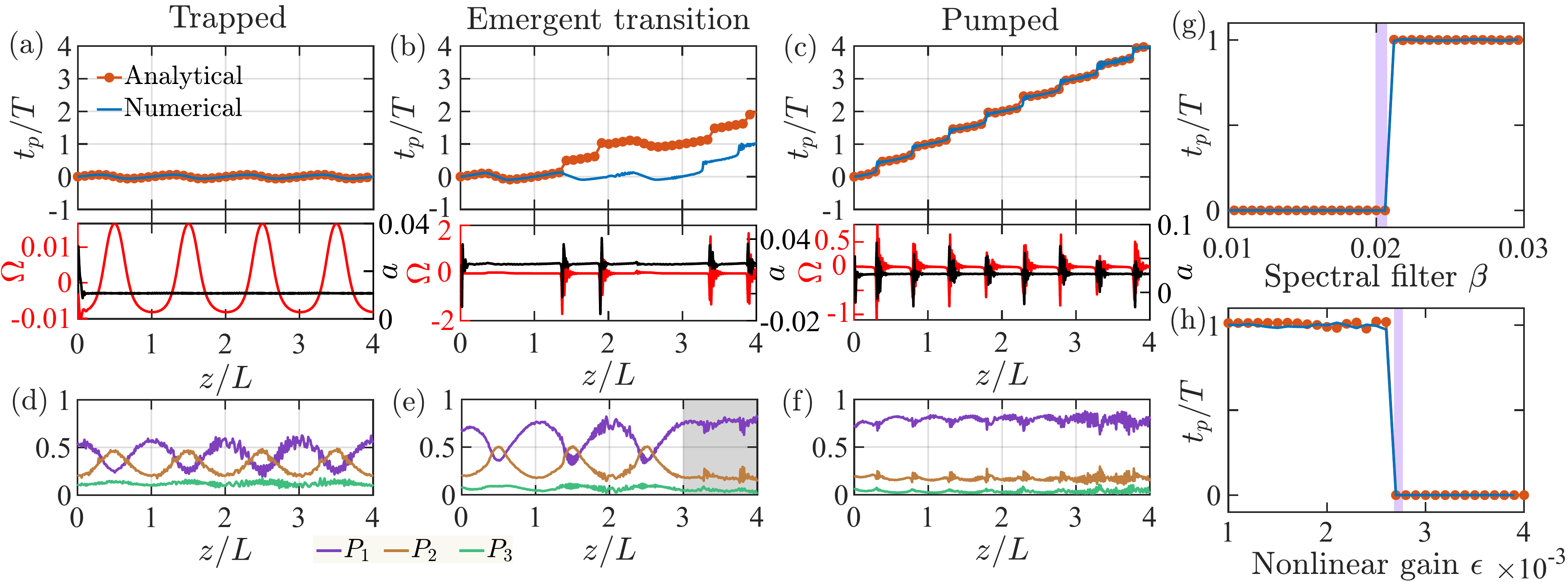}
		\caption{Topological transport and dynamically emerging topological transitions. (a)-(c) Analytical and numerical results for center-of-the-mass position $t_p(t,z)$ of the primary soliton $S_0$, when (a) $\beta=0.01$, (b) $\beta=0.02$, (c) $\beta=0.025$. Other parameters are same as Figs.~\ref{Fig2}(a)-(c). For given system parameters, same initial conditions [i.e., Eq.~(\ref{DS})] are used for numerical and analytical calculations. Bottom panels of (a)-(c) shown analytical results of the envelope parameters $\Omega(z,t)$ and $a(z,t)$ [see Eq.~(\ref{tp})]. (d)-(f) Numerical projections of the truncated soliton state onto the linear Bloch bands. Occupations $P_j$ ($j=1,2,3)$ are shown for the lowest three bands with Chern numbers $C=1,-1,1$~\cite{Supple}. In (e), the gray region depicts where the population dynamics qualitatively changes. (g)-(h) Analytical and numerical results for center-of-the-mass position of the primary soliton $S_0$ at the end of first pump cycle (i.e., $z/L=1$) as a function of (g) $\beta$ when $\delta=0.1$ and $\epsilon=0.8\epsilon_p$, and of (h) $\epsilon$ when $\beta=0.01$ and $\delta=0.1$. Purple regions denote where an emergent quantized motion occurs in successive evolutions. In all panels, the numerical results are obtained from solving CGLE, and the analytical results are obtained from solving Eq.~(\ref{Lag})~\cite{Supple}.}\label{Fig3}
	\end{figure*}

We consider the pulse propagations in nonlinear optical systems (e.g., active optical waveguides) governed by the generalized cubic CGLE~\cite{Aranson2002}, which in the dimensionless form reads as
	\begin{align}\label{CGLE}
		\!\!\!\!\!i\frac{\partial A}{\partial z}\!+\!\left[\frac{1}{2}\frac{\partial^2}{\partial t^2}\!+\!
V_{\text{ext}}+\!|A|^2\right]\!A\!=i\!\left[\beta\frac{\partial ^2}{\partial t^2}+\!\delta+\!\epsilon|A|^2\right]\!A.
	\end{align}	
Here, $A$ describes the field envelope, $z$ is the propagation distance and $t$ is the retarded
time. On the right-hand side are the dissipative terms: $\beta$ describes spectral filtering (i.e., removal of power at certain frequencies), $\delta$ is linear gain, and $\epsilon$ accounts for the nonlinear amplification. On the left-hand side are the conservative terms, including the anomalous dispersion, the focusing Kerr nonlinearity, and a space-varying temporal modulation
\begin{equation}
V_{\text{ext}}(t,z)=V_s\cos^2(2\pi t)+V_l\cos^2(\pi t-\nu z). \label{OSL}
\end{equation}
Here, a temporal modulation with the amplitude $V_l$ shifts in space at the rate $\nu$ with respect to the second modulation with the amplitude $V_s$. Equation (\ref{OSL}) realizes a Thouless pump in time domain, which is periodic both in time (the time periodicity is $T=1$) and in space (the period of one pump cycle is $L=\pi/\nu$). Without dissipation, Eq.~(\ref{CGLE}) reduces to the standard nonlinear Schr\"{o}dinger equation~\cite{Qidong2022,Fu2022,Cao2024}.

For $V_{\text{ext}}=0$, it's well known that Eq. (\ref{CGLE}) has an exact solution representing a chirped temporal soliton~\cite{Pereira1977,Akhmediev1995,Akhmediev1996}
\begin{equation}\label{DS}
    A(t,z)=\left[BC~\text{sech}\left(Bt\right)\right]^{1+id}e^{-i\omega z},
	    \end{equation}
with $d=[-3(1+2\epsilon\beta)+f_2]/{f_1}$, $f_1=2(2\beta-\epsilon)$, $f_2=\sqrt{9(1+2\epsilon\beta)^2+2f_1^2}$, $B=\sqrt{\delta/(\beta d^2+d-\beta)}$ and $C=\sqrt{3d(1+4\beta^2)/f_1}$~\cite{Supple}. Eq.~(\ref{DS}) is stable for $\delta>0$ and $\epsilon<\epsilon_p$ with $\epsilon_p=\beta(3\sqrt{1+4\beta^2}-1)/(4+18\beta^2)$. Different from conservative solitons, DS~(\ref{DS}) has fixed amplitude predetermined by system's parameters $\beta$, $\delta$ and $\epsilon$, rather than depending on the initial conditions. Figure~\ref{Fig1}(a) illustrates the peak intensity $|A|^2_\text{max}$ of Eq.~(\ref{DS}) for various spectral filtering $\beta$ and linear gain $\delta$, when $\epsilon=0.8\epsilon_p$. 

We are interested in DS dynamics once a Thouless pump is turned on (i.e., $V_\text{ext}\neq 0$). To this end, we consider an initial soliton in the form of Eq.~(\ref{DS}), and numerically solve Eq.~(\ref{CGLE}), as detailed in Supplementary Materials (SM)~\cite{Supple}. Without loss of generality, we consider a pump [Eq.~(\ref{OSL})] with $V_s=V_l=5$ and $\nu=0.1$. To elucidate the role of dissipative processes, we simulate the dynamics by varying spectral filtering $\beta$, while we fix the linear gain $\delta=0.1$ and choose $\epsilon=0.8\epsilon_p$; see Fig.~\ref{Fig2}. 

Four qualitatively distinct regimes of pumping are revealed with increased spectral filtering $\beta$ [Figs.~\ref{Fig2}(a)-(d)]: 

(I) Trapped regime [Fig.~\ref{Fig2}(a)]: For small $\beta$, we observe the initial DS ($S_0$) is trapped in time throughout entire four pump cycles. Satellite solitons are observed to emerge, similar as in pump-free case~\cite{Agrawal1991}. Once they are formed, they are trapped.

(II) Dynamically emergent phase transition regime [Fig.~\ref{Fig2}(b)]: For larger $\beta$, $S_0$ is trapped over three pump cycles, but becomes pumped at some point in the fourth cycle. As will be seen clearly in Fig.~\ref{Fig3}, the dynamical behavior of the center-of-mass position of $S_0$ changes from oscillation to unidirectional drifting in time-domain upon reaching a critical point of the propagation, thus characterizing a dynamically induced phase transition. A soliton train also emerges, which undergoes an opposite dynamical transitions from shifting to being arrested. 

(III) Pumped regime [Fig.~\ref{Fig2}(c)]: For still larger $\beta$, $S_0$ and satellite solitons synchronously shift by integer units in time. 

(IV) Collapsed regime [Fig.~\ref{Fig2}(d)]: When $\beta$ is sufficiently large, $S_0$ quickly collapses and the pulse spreads. 

To characterize the nature of soliton in the regimes (I)-(III), we monitor the peak intensity $|A(t,z)|_\text{max}^2$ and participation ratio (PR) defined as $r=1/(N_T\int|{\bar{A}(t,z)}|^4dt)$, where $\bar{A}(t,z)$ is the renormalized pulse~\cite{Supple}; see Figs.~\ref{Fig2}(e)-(g). The peak intensity characterizes the nonlinear strength of the instantaneous state, while PR quantifies its degree of localization. For sufficiently localized state, $r\approx 0$; for an extended one, the maximum possible value of $r$ is $1$. As shown, the dynamics in regimes (I)-(III) has two common features. (i) At the initial transient stage, the input soliton quickly degrades, accompanied by broadening, as the pump disrupts the original dual balances. However, subsequent oscillations of $|A(t,z)|_\text{max}^2$ and PR suggest that new double balances are dynamically formed. (ii) Beyond $z/L\approx 2$, a steep increase in PR is observed, albeit its peak value remains small, aligning with the birth of satellite solitons. The generation of a train of solitons has been known in pump-free case~\cite{Agrawal1991} and is related to gain medium: The pump-induced fluctuations are amplified by gain, and are shaped into chirped solitons through spectral filtering. The amplification continues until the soliton reaches the amplitude determined by system parameters, as illustrated by Fig.~\ref{Fig2}(h) for the satellite soliton $S_1$.

On the other hand, Figs.~\ref{Fig2}(e)-(g) show the (overall) pulse power decreases with $\beta$, consistent with Fig.~\ref{Fig1}(a2). Moreover, in the trapped and pumped regimes [Figs.~\ref{Fig2}(d) and (g)], the average peak intensity is roughly constant after the initial transient stage. In contrast, in regime (II) [Fig.~\ref{Fig2}(f)], there is a pronounced drop of $|A(t,z)|_\text{max}^2$ at $z/L\gtrsim 3$, where the transition of $S_0$ from arrested to pumped is observed [Fig.~\ref{Fig2}(b)]. Such a drop cannot occur in conservative systems where the evolution is unitary.

To make explicit the distinct topological behaviors in (I)-(III), we numerically calculate the center-of-mass position of $S_0$ [blue curves in Figs.~\ref{Fig3}(a)-(c)], i.e., 
\begin{equation}
\!\!\!\!\!\!\!\!\!	t_{p}(z)=\!\!\int_{t\in ts} t\left|{A}_p(t,z)\right|^2dt,  \label{sigma}
\end{equation}
where $t\in{t}_\text{s}$ is such that it contains solely $S_0$, and $A_p$ denotes the normalized truncated pulse~\cite{Supple} within $t\in t_\text{s}$. For $\beta=0.01$ [Fig.~\ref{Fig3}(a)], the path of $S_0$ oscillates around its initial position, whereas for $\beta=0.025$ [Fig.~\ref{Fig3}(c)], $S_0$ shifts by $4$ units at the end of four cycles. Although the trapped-to-pumped transition has been known in conservative solitons, it is here driven by dissipation; this is conclusively summarized in Figs.~\ref{Fig3}(g) and (h), where we plot center-of-the-mass position at the end of first pump cycle as a function of $\beta$ [Fig.~\ref{Fig3}(g)] and $\epsilon$ [Fig.~\ref{Fig3}(h)], respectively. Unique for dissipative pumping is Fig.~\ref{Fig3}(b), where the trapped-to-pumped transition occurs during the course of propagation, even though the system parameters are fixed.

Thus, Figs.~\ref{Fig2} and \ref{Fig3} indicate three major differences of dissipative nonlinear pumping compared to conservative cases: (1) The dynamics is nonunitary. (2) The trapped or pumped regimes are determined by dissipative mechanisms. (3) The dissipative dynamics itself can induce an emergent topological phase transition.  

To get more physical insights into the mechanisms under dissipative nonlinear pumping, we carry out an analytical study based on the variational Lagrangian approach for dissipative systems~\cite{Kivshar1995,Filho2001,Skarka2006,Roy2013,Sahoo2017,Tusnin2020}. The starting point is a variational solution to CGLE (\ref{CGLE}), i.e., 
\begin{equation}
\!\!\!\!\!A(t,z)\!=\!A_0\left\{\text{sech}[\eta(t-t_p)]\right\}^{1+ia}e^{i\left[\phi-\Omega(t-t_p)\right]}.\label{TrialW}
\end{equation}
This ansatz assumes the same functional form as Eq.~(\ref{DS}), except that all the profile parameters now evolve with propagation, i.e., $\{q_j(t,z)\}_{j=1,...,6}\equiv A_0(t,z), a(t,z), \eta(t,z), t_p(t,z), \Omega(t,z), \phi(t,z)$ denote the six variational parameters associated with the amplitude, width, center-of-mass position and frequency, and phase, respectively. In the limit $\nu\rightarrow 0$, finding the variational solution (\ref{TrialW}) from the Lagrangian approach can be understood as finding the instantaneous nonlinear eigensolution to CGLE~(\ref{CGLE}) at each $z$.  

Following Refs.~\cite{Kivshar1995,Filho2001,Skarka2006,Roy2013,Sahoo2017,Tusnin2020}, we derive~\cite{Supple} the Euler-Lagrange equations for $q_j(t,z)$ ($j=1,...,6$), i.e.,
\begin{equation}\label{Lag}
\!\!\!\!\!\frac{\partial L}{\partial q_j}-\!\frac{d}{dz}\frac{\partial L}{\partial \dot{q}_j}\!=\!-2\!\int_{-\infty}^{+\infty}\!\!\!\!\!\!\!\!\text{Im}{\left[\frac{\partial A^*}{\partial q_j}(\beta\frac{\partial^2}{\partial t^2}+\!\delta+\epsilon|A|^2)A\right]\!dt}.
\end{equation}
Here, the Lagrangian $L=\int_{-\infty}^{+\infty}[iA^\star\partial_ zA-(\frac{1}{2}\left|{\partial_t A}\right|^2-\frac{1}{2}\left|A\right|^4-V_{\text{ext}}\left|A\right|^2)]dt$. Straightforward calculations~\cite{Supple} yield $L=L_0+L_1+L_2$, where $L_1=\frac{2}{3\eta}A_0^4$ arises from the Kerr nonlinearity, $L_2=\frac{A_0^2\pi^2}{\eta^2}[2V_s\cos(4\pi t_p)\text{csch}(\frac{2\pi^2}{\eta})+V_l\cos(2\pi t_p-2\nu z)\text{csch}(\frac{\pi^2}{\eta})]+\frac{A_0^2}{\eta}(V_s+V_l)$ accounts for the pump, and $L_0$ contains all remaining terms associated with conservative processes. Note that the width $\eta$ has a lower bound due to the finite gain bandwidth~\cite{Agrawal1991}: when the pulse becomes so short that its spectrum exceeds the gain bandwidth, spectral wings are amplified less than the central peak. Importantly, the right side of Eq.~ (\ref{Lag}) are the dissipative forces acting on DS in time-domain due to energy exchanges with its environment. 

We shall present only the equations for center-of-the-mass position $t_p$ and frequency $\Omega$, delegating the remaining four equations in SM~\cite{Supple}, i.e.,
\begin{eqnarray}
\!\!\!\!\!\!\!\!\!\!	\frac{dt_p(z)}{dz}&=&-\Omega(z)-2\beta\Omega(z)a(z), \label{tp}\\
\!\!\!\!\!\!\!\!\!\!	\frac{d\Omega(z)}{dz}&=&-\frac{4}{3}\beta\Omega(z)\eta^2(z)[1+a^2(z)]+f_s(\eta)V_s\sin(4\pi t_p)\nonumber\\
&+&f_l(\eta)V_l \sin(2\pi t_p-2\nu z),\label{omega}
\end{eqnarray}
where $f_s=\frac{4\pi^3}{\eta(z)}\text{csch}\left(\frac{2\pi^2}{\eta(z)}\right)$ and $f_l=\frac{\pi^3}{\eta(z)}\text{csch}\left(\frac{\pi^2}{\eta(z)}\right)$. As shown in Figs.~\ref{Fig3}(a)-(c) and (g)-(h), the analytical results of $t_p(z)$ (red curves) not only quantitatively agree with the numerical results in trapped and pumped regimes, it also predicts the occurrence of dynamical phase transition in the regime (II). This indicate that Eqs.~(\ref{TrialW}) and (\ref{Lag}) capture the essential physics of dissipative nonlinear pumping, and also that the dynamical transition is not due to the generation of satellite solitons. Equations~(\ref{tp}) and (\ref{omega}) reveal two types of forces acting on center-of-the-mass of a DS: a conservative force from the pump and a frictional force $\sim\beta\Omega$ depending on the soliton profile. The lower panels of Figs.~\ref{Fig3}(a)-(c) depict analytical results of $\Omega(t)$ and $a(t)$ for various $\beta$. There, the envelope in the trapped regime exhibits smooth oscillations with the same period as the pump, in sharp contrast to the periodic spikes in the pumped regime. 

While dissipative nonlinear dynamics is significantly different from the conservative case, the DS transport can still be understood from the topology of the linear Bloch bands [Figs.~\ref{Fig3}(d)-(f)]. The linear part of CGLE (\ref{CGLE}) is given by $i\partial_z{A}=H_0A$ with a non-Hermitian linear Hamiltonian
\begin{equation}
H_0=-\frac{1}{2}(1-2i\beta)\frac{\partial^2}{\partial t^2}+V_\text{ext}(t,z)+i\delta.\label{H0}
\end{equation}
Apart from the gain $\sim \delta$, Equation (\ref{H0}) effectively describes Thouless pumping of a particle with complex mass $1/m^*=1-2i\beta$. Given $\beta\ll 1$ here, we expect the non-Hermitian Bloch bands in time-space domain and thus their topological characterization are similar as Hermitian ones~\cite{Supple}; we refer to SM for calculations of non-Hermitian bands. In Figs.~\ref{Fig3}(d)-(f), we show numerical projections of the truncated primary soliton state $A_0(t,z)$ onto the lowest three bands with Chern numbers $C=\{1, -1, 1\}$~\cite{Supple}. For $\beta=0.02$ [Fig.~\ref{Fig3}(f)], mainly the lowest band with $C=1$ is occupied, explaining the unidirectional shift by one unit in each cycle. For $\beta=0.01$ [Fig.~\ref{Fig3}(d)], both bands are occupied, hence the net displacement essentially follows the zero total Chern number of the two bands. Particularly intriguing is Fig.~\ref{Fig3}(e), where the occupation behavior dynamically changes from Rabi oscillations between two bands to essentially occupying only the lowest band, manifesting the emergent topological transition observed in Fig.~\ref{Fig2}(b).

Summarizing, we present the first study of dissipative nonlinear Thouless pumping. We find that the DS's transport in systems with gain and loss can be quantized. Here, dissipation plays a unique, pivotal role in quantizing soliton's motion, both through the double-balance mechanism and through pulse shaping in nonunitary evolutions. Hence, nonlinear topological transition not only occurs when modulating dissipative parameters, it can be emergent in dynamics. The predicted phenomena generically hold for various system parameters~\cite{Supple}, and is experimentally feasible with state-of-the-art techniques in synthetic materials such as the Kerr nonlinear optical micro-resonators~\cite{Anderson2023}. Given that the CGLE is central in describing dissipative nonlinear phenomena in a wide range of settings, from optics and photonics to fluid dynamics and condensed matter, our work opens broad prospect for exploring those nonlinear topological phenomena unique to dissipative settings.

\textit{Acknowledgements} -- We thank Qidong Fu, Nan Li, Fangwei Ye, Qi Zhang, and Biao Wu for stimulating discussions and useful help. This work was supported by the National Natural Science Foundation of China (Grants No. 12074344, No. 12374246) and the Zhejiang Provincial Natural Science Foundation (Grant No. LZ21A040001). Y.H. acknowledges support by Beijing National Laboratory for Condensed Matter Physics (Grant No. 2023BNLCMPKF001).

\bibliography{Reference}

\end{document}


\title{Supplementary Materials for ``Dissipative Nonlinear Thouless Pumping of Temporal Solitons"}
	
	\author{Xuzhen Cao}
	\affiliation{State Key Laboratory of Quantum Optics and Quantum Optics Devices, Institute of Laser Spectroscopy, Shanxi University, Taiyuan 030006, China}
	\affiliation{Collaborative Innovation Center of Extreme Optics, Shanxi University, Taiyuan 030006, China}

	\author{Chunyu Jia}
	\affiliation{Department of Physics, Zhejiang Normal University, Jinhua 321004, China}
	
	\author{Ying Hu}
	\thanks{huying@sxu.edu.cn}
	\affiliation{State Key Laboratory of Quantum Optics and Quantum Optics Devices, Institute of Laser Spectroscopy, Shanxi University, Taiyuan 030006, China}
	\affiliation{Collaborative Innovation Center of Extreme Optics, Shanxi University, Taiyuan 030006, China}

	\author{Zhaoxin Liang}
	\thanks{zhxliang@zjnu.edu.cn}
	\affiliation{Department of Physics, Zhejiang Normal University, Jinhua 321004, China}
	
	\maketitle
	
The Supplemental Materials are structured as follows. In Sec.~\ref{AppendixA}, we give the detailed derivation of the exact soliton solution in Eq.~(3) of the main text. In Sec.~\ref{AppendixB}, we describe our numerical scheme for solving the CGLE~(1) of the main text, and other details relevant for obtaining the results in Figs.~2 and 3. In Sec.~\ref{AppendixC}, we present the variational Lagrangian approach and derive the full set of equations of motion of the six variational parameters contained in the variational ansatz [Eq.~(5)] of the main text. In Sec.~\ref{AppendixD}, we follow the standard procedures to calculate Chern numbers of the non-Hermitian linear bands. In Sec.~\ref{AppendixE}, we show generality of our key results under various linear gain $\delta$, nonlinear gain $\epsilon$, $dt$, and system size $N_T$.

\section{Dissipative Solitary solution of CGLE without Thouless pump}\label{AppendixA}

Following Refs.~\cite{Akhmediev1995,Akhmediev1996},  we present detailed derivations of the exact soliton solution [Eq.~(3) in main text] in this section. For the sake of self-consistence, we rewrite the cubic CGLE [Eq.~(1) in main text] as follows,

\begin{align}\label{Eq:CGLE_Appendix}
	i\frac{\partial A}{\partial z}+\left(\frac{1}{2}-i\beta\right)\frac{\partial^2A}{\partial t^2}+|A|^2A=\left[i\delta+i\epsilon|A|^2\right]A.
\end{align}
Here, we are interested in the DS solution of Eq. (\ref{Eq:CGLE_Appendix}) with the following form
\begin{equation}\label{Eq:FormalSolution}
	A(t,z)=a(t)\exp(id\ln[a(t)]-i\omega z),
\end{equation}
where $a$ is the amplitude, $d$ is the chirp parameter, and $\omega$ is the speed of propagation along $z$. To determine the concrete form of these three parameters, we plug Eq.~(\ref{Eq:FormalSolution}) into Eq.~(\ref{Eq:CGLE_Appendix}) and separate them into real and imaginary terms
\begin{eqnarray}
	\omega a+\left(\frac{1}{2}+d\beta\right)a''+\left(\frac{d\beta}{a}-\frac{d^2}{2a}\right)(a')^2+a^3=0, \label{a1}\\
	-\delta a+\left(\frac{d}{2}-\beta\right)a''+\left(\frac{d}{2a}+\frac{d^2\beta}{a}\right)(a')^2-\epsilon a^3=0,\label{a2}
\end{eqnarray}
with the notations of $a'=da/dt$ and $a''=d^2a/dt^2$. Elimination of $a'$ from Eqs.~(\ref{a1}) and (\ref{a2}) yields
\begin{equation}\label{Eq:a2d}
	c_1\frac{a''}{a}+c_2a^2+c_3=0,
\end{equation}
with $c_1=\frac{d}{4}(1+d^2)(1+4\beta^2)$, $c_2=\frac{1}{2}(\frac{d}{2}+\beta d^2+\epsilon d\beta-\frac{\epsilon d^2}{2})$, and $c_3=\frac{\omega d}{2}(1+2d\beta)+\delta(d\beta-\frac{d^2}{2})$. Using the identity $a''=\partial_t (a')^2/(2a')$, Eq.~(\ref{Eq:a2d}) can further be derived into the following form 
\begin{equation}\label{Eq:ea1}
	c_1\frac{(a')^2}{a^2}+{c}_2a^2+{c}_3=0.
\end{equation}
 Alternatively, one can eliminate $a''$ from Eqs.~(\ref{a1}) and (\ref{a2}) to obtain 
\begin{eqnarray}\label{Eq:ea2}
	\tilde{c}_1\frac{(a')^2}{a^2}+\tilde{c}_2a^2+\tilde{c}_3=0,
\end{eqnarray}
with the different set of coefficients $\tilde{c}_1=\frac{d}{4}(1+d^2)(1+4\beta^2)$, $\tilde{c}_2=\beta-\frac{d}{2}-\frac{\epsilon}{2}-\epsilon d\beta$, and $\tilde{c}_3=-\delta(\frac{1}{2}+d\beta)-\frac{\omega d}{2}+\omega\beta$. Since Eqs.~(\ref{Eq:ea1}) and (\ref{Eq:ea2}) must be equal with each other by equating the coefficients $c_j=\tilde{c}_j$ ($j=1,2,3$), we find
\begin{eqnarray}
	d_\pm&=&\frac{3+6\beta\epsilon\pm\sqrt{9+32\beta^2+4\beta\epsilon+8\epsilon^2+36\beta^2\epsilon^2}}{2(\epsilon-2\beta)},\label{eq:d}\\
	\omega&=&-\frac{\delta(1-d^2+4d\beta)}{2(d-\beta+d^2\beta)}\label{Eq:dandw}.
\end{eqnarray}
Finally, substituting Eqs.~(\ref{eq:d}) and (\ref{Eq:dandw}) back into Eq.~(\ref{Eq:ea2}), and having in mind that the solution (\ref{Eq:FormalSolution}) should vanish at infinity, we find 
\begin{align}
	a_\pm=\pm BC\text{sech}(Bt),
\end{align}
with the coefficients
\begin{align}
	B=\sqrt{\frac{\delta}{\beta d^2+d-\beta}},~~ C=\sqrt{\frac{3d(1+4\beta^2)}{2(2\beta-\epsilon)}}\label{AppAB}.
\end{align}
By plugging Eqs. (\ref{AppAB}), and (\ref{eq:d}) back  into Eq. (\ref{Eq:FormalSolution}), we obtain the exact expression of DS of CGLE (\ref{Eq:CGLE_Appendix}) corresponding to Eq. (3) of the main text as follows
\begin{equation}\label{AppDS}
    A(t,z)=\left[BC~\text{sech}\left(Bt\right)\right]^{1+id}e^{-i\omega z},
	    \end{equation}
with $d=[-3(1+2\epsilon\beta)+C_2]/{C_1}$ with $C_1=2(2\beta-\epsilon)$ and $C_2=\sqrt{9(1+2\epsilon\beta)^2+2C_1^2}$, $B=\sqrt{\delta/(\beta d^2+d-\beta)}$ and $C=\sqrt{3d(1+4\beta^2)/C_1}$. Note that the resulting form of soliton solution (\ref{Eq:FormalSolution}) is also known as the solution of Pereira and Stenllo~\cite{Pereira1977}.

In Fig.~1(a2) of the main text, we have shown the peak power of the solitary pulse as a function of $\delta$ and $\beta$ under the condition $\epsilon=0.8\epsilon_p$. In the supplementary Fig.~\ref{Fig:Suppfig1}, we fix $\delta=0.1$ and show the peak intensity for various $\epsilon$ and $\beta$. 

\begin{figure}[t]
	\centering
	\includegraphics[width=0.5\columnwidth]{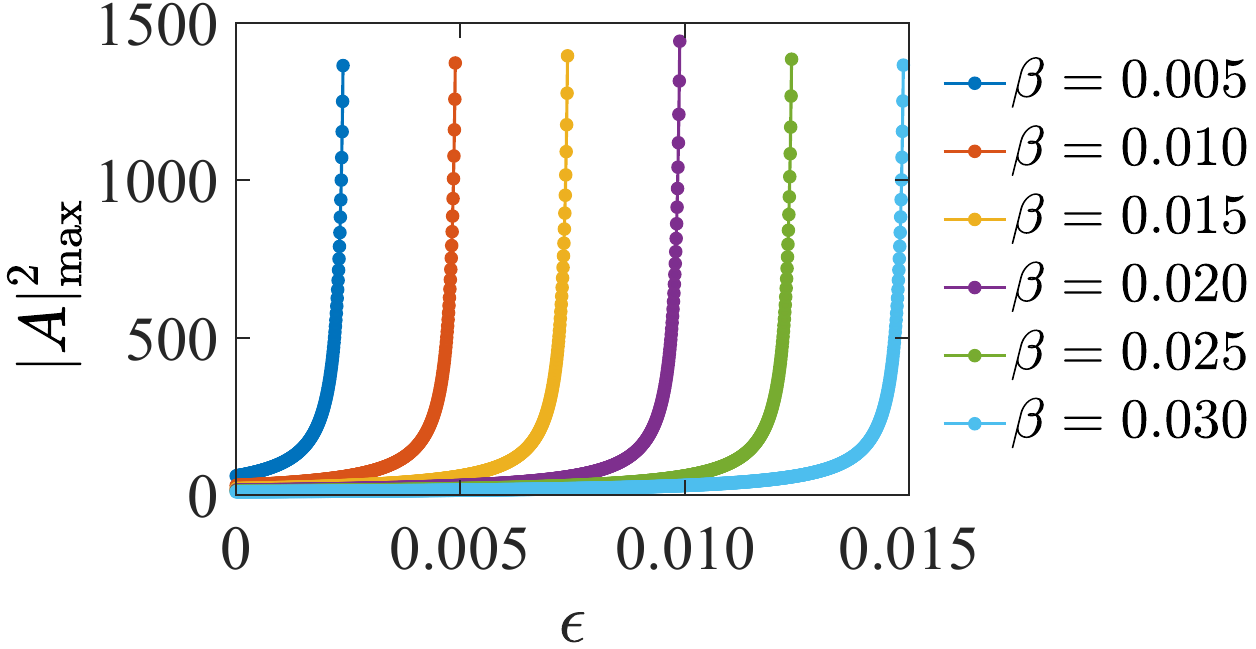}
	\caption{Peak intensity of the exact soliton solution (Eq.~3 of main text) as a function of nonlinear gain $\epsilon$ for various spectral filtering $\beta$, when linear gain $\delta=0.1$.}\label{Fig:Suppfig1}
\end{figure}

\section{Numerical simulation of CGLE}\label{AppendixB}

In this section, we present details on our numerical methods, which include (i) the solution of the CGLE using the second-order split-step fast Fourier algorithm (see e. g., Ref.~\cite{Feit1979} and the references therein), and (ii) the extraction of the primary soliton, as relevant for Eq.~3 and the numerical results in Fig.~3 of the main text.

We begin with describing our numerical schemes to solve  the CGLE. To begin with, we rewrite the CGLE~(1) of main text as 
\begin{align}
	i\frac{\partial}{\partial z}A=(O_1+O_2)A,
\end{align}
where the two operators are
\begin{eqnarray}
	O_1&=&\left(i\beta-\frac{1}{2}\right)\frac{\partial^2}{\partial t^2},\nonumber\\
	O_2&=&i\delta +(i\epsilon-1)|A|^2+V_{\text{ext}}.
\end{eqnarray}
Note that $[O_1,O_2]\neq0$. To solve the time evolution, we implement the second-order splitting scheme
\begin{align}\label{Eq:SSFM1}
	A(t,z+dz)=U_2U_1U_2A(t,z),
\end{align}
where $U_1=\exp(-iO_1dz)$ and $U_2=\exp(-iO_2dz/2)$
with a sufficiently small spatial step $dz=10^{-5}$. We use the fast Fourier transform (FFT) algorithm to compute Eq.~(\ref{Eq:SSFM1}). Specifically, let $\mathcal{F}$ denotes the Fourier transform and $\mathcal{F}^{-1}$ denotes the inverse Fourier transform, we have
\begin{eqnarray}
A(t,z+dz)=U_2\mathcal{F}^{-1}\left\{\exp\left[\left(-\frac{i}{2}-\beta\right)k^2 dz\right]\mathcal{F}[U_2A(t,z)]\right\}.
\end{eqnarray}
Finally, we choose $dt=2.5\times10^{-3}$ in the time discretization of the CGLE. 

To benchmark our numerics, we make use of the exact solution~(\ref{Eq:FormalSolution}) in free space, and compare the numerically evolved soliton in the absence of pump with the exact solution by calculating their intensity difference $D=\left||A_{\text{exa}}|^2-|A_{\text{num}}|^2\right|$. The results are shown in the supplementary Fig.~\ref{FigSupp2} for various system parameters, which indicate sufficient accuracies of our numerics. 

Finally, we describe how we obtain the truncated soliton state $A_0(t,z)$ described in Eq.~4 and Fig.~3 of the main text. In Fig.~3, we calculated the center-of-the-mass position of only the primary soliton $S_0$ as well as its projection onto the linear Bloch states. This is possible when the soliton remains well localized in the relevant parameter regimes, i.e., its width is much smaller compared to $T$, which is the temporal periodicity of the Thouless pump. In our case, we have $T=1$. Numerically, to isolate the primary soliton from the multisoliton state, we retain only the pulse distribution $A(t,z)$ within $t\in [t_p-0.5T,t_p+0.5T]$, where $t_p$ is the center-of-the-mass-position of the primary soliton, while ignoring the pulses in the regions $t< t_p-0.5T$ and $t>t_p+0.5T$. This truncated pulse is then renormalized to yield $A_p(t,z)$ used in Eq.~4 of the main text.
\begin{figure}[t]
	\centering
	\includegraphics[width=0.7\columnwidth]{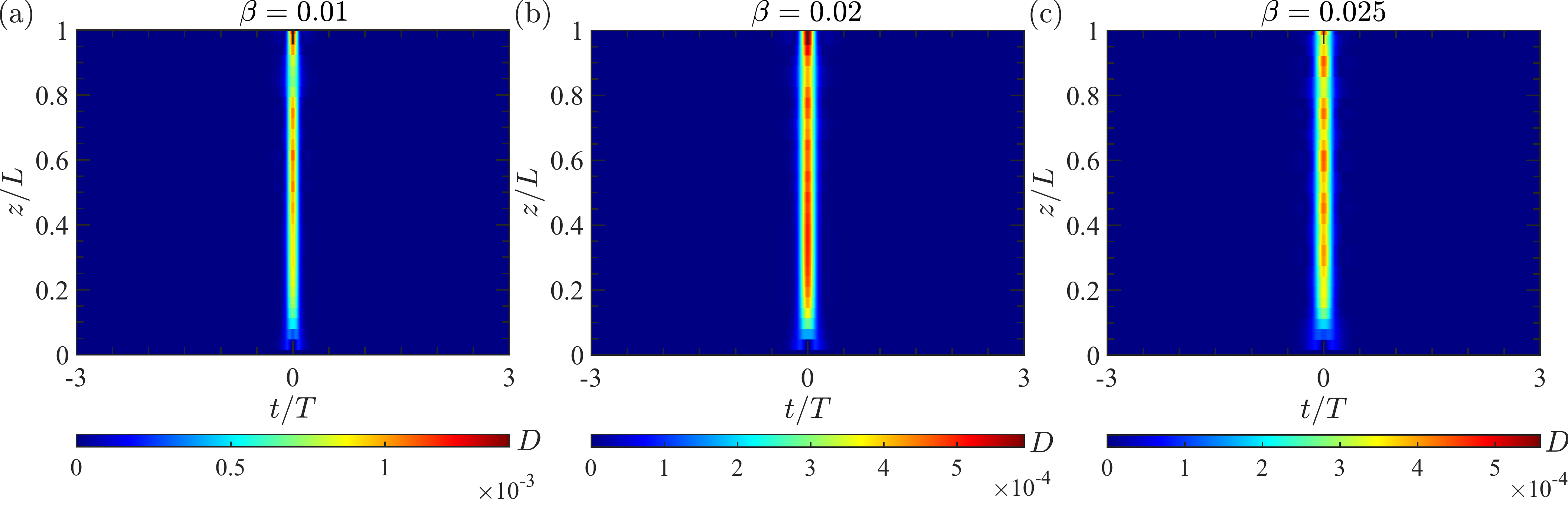}
	\caption{Comparisons between the exact solution [Eq.~(3) in main text] and the numerically obtained free-space soliton solution. The difference is measured through the quantity $D=\left||A_{\text{exa}}|^2-|A_{\text{num}}|^2\right|$, where $A_{\text{exa}}$ and $A_{\text{num}}$ denote the amplitude of the exact solution and numerical solution, respectively. (a) $\delta=0.1$, $\beta=0.01$, $\epsilon=0.8\epsilon_p\approx0.004$, (b) $\delta=0.1$, $\beta=0.02$, $\epsilon=0.8\epsilon_p\approx0.008$, and (c) $\delta=0.1$, $\beta=0.025$, $\epsilon=0.8\epsilon_p\approx0.01$, which correspond to Fig.~2(a), Fig.~2(b), and Fig.~2(c) in main text, respectively.}\label{FigSupp2}
\end{figure}

\begin{figure}[t]
	\centering
	\includegraphics[width=0.8\columnwidth]{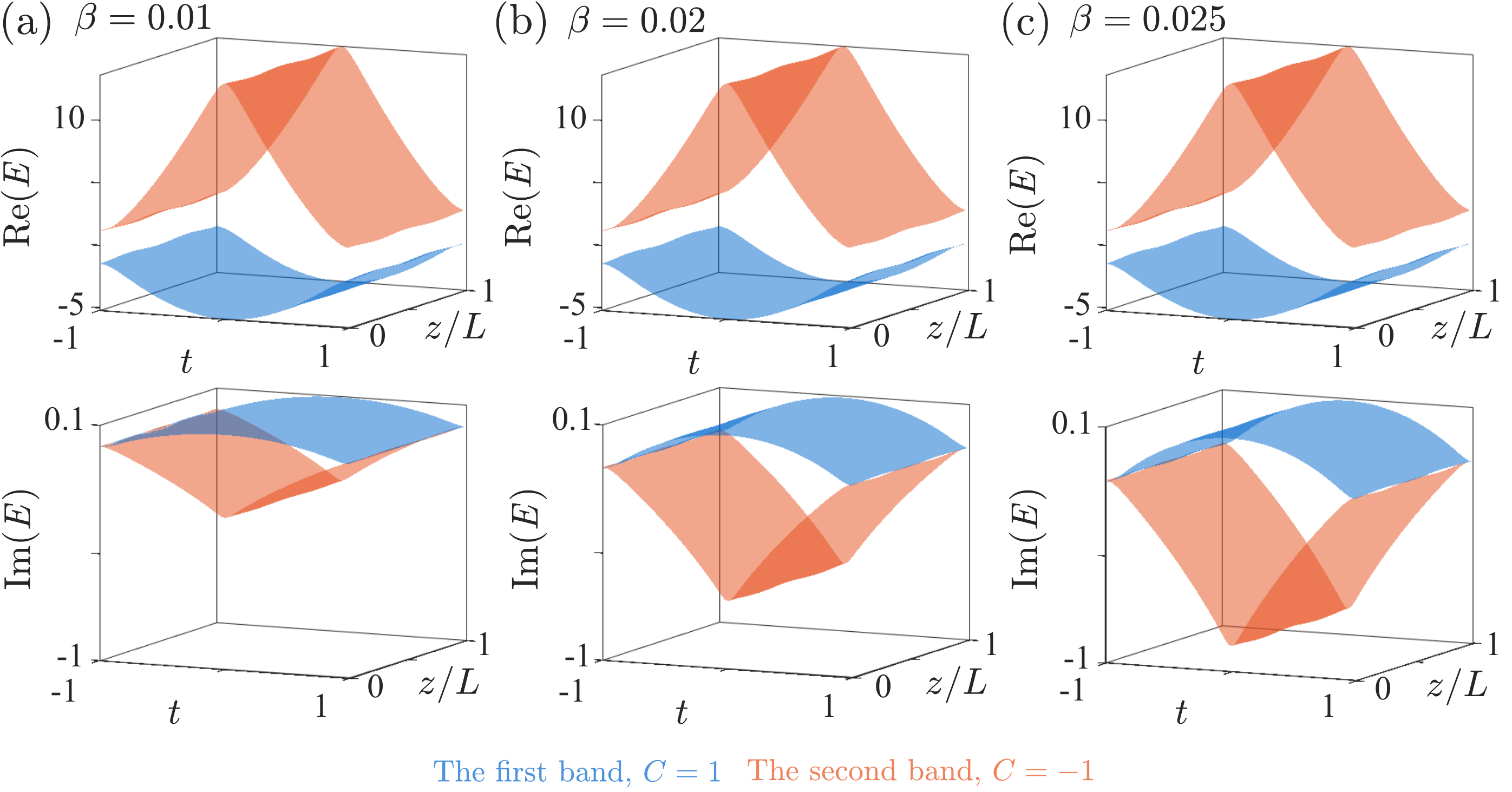}
	\caption{Real and imaginary parts of the complex eigenvalues of the two lowest non-Hermitian Bloch bands (i.e., ordered according to real components of the complex eigenvalues). }\label{FigSupp3}
\end{figure}

\section{Lagrangian variational approach for dissipative systems}\label{AppendixC}

Following Ref.~\cite{Kivshar1995,Filho2001,Skarka2006,Roy2013,Sahoo2017,Tusnin2020}, we derive the Euler-Lagrange equations for the six variational parameters in Eqs.~(5) of the main text using the Lagrangian variational approach for a dissipative nonlinear system. 

As a first step, we rewrite  CGLE (\ref{Eq:CGLE_Appendix}) into the form of a perturbed nonlinear Schr\"odinger equation
\begin{align}
	\!\!i\frac{\partial A}{\partial z}+\frac{1}{2}\frac{\partial^2A}{\partial t^2}\!+\!|A|^2A-V_{\text{ext}}A=ip(A),\label{AppCCGLE}
\end{align}
where $ip(A)$ is treated as the perturbation, reading,
\begin{align}
	p(A)=\left(\beta\frac{\partial^2}{\partial t^2}\!+\delta+\epsilon|A|^2\right)A.
\end{align}
When dissipative perturbation vanishes (i.e., $p=0$), the system is reduced to a conservative system. 

Next, we follow the Lagrangian variational approach in Ref.~\cite{Kivshar1995,Filho2001,Skarka2006,Roy2013,Sahoo2017,Tusnin2020} to solve CGLE (\ref{AppCCGLE}) analytically. We assume a trial solution of CGLE (\ref{AppCCGLE}) with the following form
\begin{align}\label{ansatz}
	A(z,t)=A_0 \{\text{sech}[\eta(t-t_p)]\}^{1+ia}\exp[i(\phi-\Omega(t-t_p))]
\end{align}
where $A_0$, $a$, $\eta$, $t_p$, $\phi$ and $\Omega$ are the six variational parameters to be determined. We denote 
\begin{equation}
N=\int_{-\infty}^{+\infty}|A|^2dt=2A_0^2/\eta.
\end{equation}

When $p=0$, Equation~(\ref{AppCCGLE}) describes a conservative system. The corresponding Lagrangian $L$ can be written as
\begin{align}\label{Lagrangian1}
	L=\frac{i}{2}\int_{-\infty}^{+\infty}\left(A^*\frac{\partial A}{\partial z}-A\frac{\partial A^*}{\partial z}\right)dt-E,
\end{align}
where $E$ is the system energy given by
\begin{eqnarray}
	E=\frac{1}{2}\int_{-\infty}^{+\infty}\left|\frac{\partial A}{\partial t}\right|^2dt-\frac{1}{2}\int_{-\infty}^{+\infty}\left|A\right|^4dt+\int_{-\infty}^{+\infty}V_{\text{ext}}(t,z)\left|A\right|^2dt.
\end{eqnarray}
Inserting ansatz Eq.~(\ref{ansatz}) into Eq.~(\ref{Lagrangian1}), after straightforward calculations, we obtain
\begin{eqnarray}
	L&=&-\frac{A_0^2}{\eta}\frac{da}{dz}[\ln(4)-2]-\frac{2A_0^2}{\eta}\left(\frac{d\phi}{dz}+\Omega\frac{dt_p}{dz}\right)+\frac{aA_0^2}{\eta^2}\frac{d\eta}{dz}-\frac{A_0^2}{\eta}\Omega^2-\frac{\eta A_0^2}{3}(1+a^2)\nonumber\\
	&&+\frac{2}{3}\frac{A_0^4}{\eta}+\frac{V_s A_0^2}{\eta^2}\left[2\pi^2\cos\left(4\pi t_p\right)\text{csch}\left(\frac{2\pi^2}{\eta}\right)+\eta\right]+\frac{V_lA_0^2}{\eta^2}\left[\pi^2\cos\left(2\nu z-2\pi t_p\right)\text{csch}\left(\frac{\pi^2}{\eta}\right)+\eta\right].	\label{AppCL}
\end{eqnarray}
We note that there is a lower bound on $\eta$ due to the gain bandwidth. 

When $p\neq0$,  it is necessary to introduce the generalized friction term into the Euler-Lagrange equation. Directly following Ref.~\cite{Kivshar1995,Filho2001,Skarka2006,Roy2013,Sahoo2017,Tusnin2020}, the Euler-Lagrange equation in a dissipative nonlinear system can be written as
\begin{align}
	\frac{\partial L}{\partial q_j}-\frac{d}{dz}\frac{\partial L}{\partial \dot{q}_j}=\int_{-\infty}^{+\infty}i\left(p\frac{\partial A^*}{\partial q_j}-p^*\frac{\partial A}{\partial q_j}\right)dt. \label{AppCEL}
\end{align}
Plugging Eq.~(\ref{AppCL}) into Eq.~(\ref{AppCEL}), and using $N=\int_{-\infty}^{+\infty}|A|^2dt=2A_0^2/\eta$, after some tedious calculations, we derive the Euler-Lagrange equations for the variational parameters $q_j= t_p, \Omega,N, a, \eta, \phi$ as
\begin{eqnarray}
\frac{dt_p}{dz}&=&-\Omega(1+2a\beta),\label{AppCt}\\	
\frac{d\Omega}{dz}&=&-\frac{4}{3}\beta\Omega\eta^2(1+a^2)+\frac{4\pi^3V_s}{\eta}\sin(4\pi t_p)\text{csch}\left(\frac{2\pi^2}{\eta}\right)-\frac{\pi^3V_l}{\eta}\sin(2\nu z-2\pi t_p)\text{csch}\left(\frac{\pi^2}{\eta}\right),\label{AppCOmega}\\
\frac{dN}{dz}&=&\frac{4A_0^2}{\eta}(\delta-\beta\Omega^2)-\frac{4}{3}\beta A_0^2\eta(1+a^2)+\frac{8\epsilon A_0^4}{3\eta},\label{N}\\
\frac{d\eta}{dz}&=&\frac{2}{3}\eta^3a-\frac{\eta^2}{2A_0^2}\frac{dN}{dz}[\ln(4)-2]+2\eta(\delta-\beta\eta^2-\beta\Omega^2)[\ln(4)-2]+\frac{4}{9}\eta\epsilon A_0^2[\ln(64)-5]\nonumber\\
&+&\frac{2}{9}\beta\eta^3(2-a^2)[\ln(64)-8],\label{eta}\\
\frac{da}{dz}&=&2a\delta+\frac{2}{3}a\epsilon A_0^2-2a\beta\Omega^2-\frac{4}{3}a\beta\eta^2(1+a^2)-\frac{2}{3}\eta^2(1+a^2)+\frac{2}{3}A_0^2-\frac{a}{N}\frac{dN}{dz}\nonumber\\
	&-&\frac{2\pi^2V_s}{\eta}\cos(4\pi t_p)\text{csch}\left(\frac{2\pi^2}{\eta}\right)-\frac{\pi^2V_l}{\eta}\cos(2\nu z-2\pi t_p)\text{csch}\left(\frac{\pi^2}{\eta}\right)\nonumber\\
	&+&\frac{4\pi^4V_s}{\eta^2}\cos(4\pi t_p)\coth\left(\frac{2\pi^2}{\eta}\right)\text{csch}\left(\frac{2\pi^2}{\eta}\right)+\frac{\pi^4V_l}{\eta^2}\cos(2\nu z-2\pi t_p)\coth\left(\frac{\pi^2}{\eta}\right)\text{csch}\left(\frac{\pi^2}{\eta}\right),\label{a}\\
\frac{d\phi}{dz}&=&\frac{2}{3}A_0^2-\frac{1}{6}\eta^2(1+a^2)-\frac{1}{2}\Omega^2-\frac{1}{2}\frac{da}{dz}[\ln(4)-2]+\frac{a}{2\eta}\frac{d\eta}{dz}-\Omega\frac{dt_p}{dz}\nonumber\\
	&+&\frac{V_s}{2\eta}\left[2\pi^2\cos(4\pi t_p)\text{csch}\left(\frac{2\pi^2}{\eta}\right)+\eta\right]+\frac{V_l}{2\eta}\left[\pi^2\cos(2\nu z-2\pi t_p)\text{csch}\left(\frac{\pi^2}{\eta}\right)+\eta\right]. \label{phi}
\end{eqnarray}
Equations (\ref{AppCt}) and (\ref{AppCOmega}) correspond to Eqs.~(7) and (8) of the main text. 

The six equations (\ref{AppCt})-(\ref{phi}) are supplemented by the initial conditions. In our work, the initial soliton state is given by Eq.~(3) of the main text, which is predetermined for given dissipative parameters $\delta$, $\beta$ and $\epsilon$. Specifically, we use the initial conditions: $N(0)=2|A|^2_{\text{max}}/\eta(0)$, $t_p(0)=0$, $\Omega(0)=0$, $\eta(0)=\sqrt{\delta/(\beta d^2+d-\beta)}$, $a(0)=d$ and $\phi(0)=0$. 
\begin{figure}[t]
	\centering
	\includegraphics[width=0.8\columnwidth]{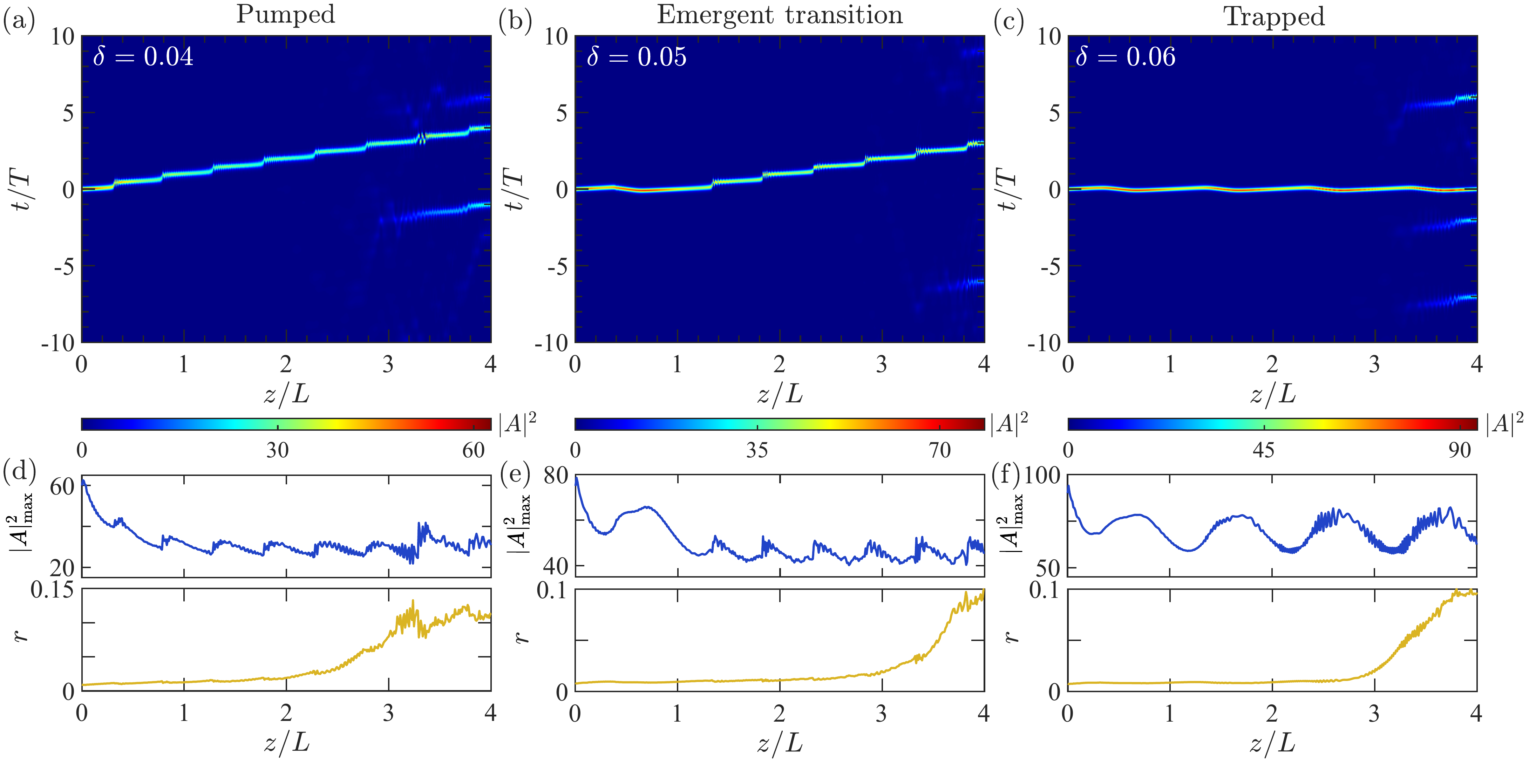}
	\caption{Pumped, emergent transition, and trapped regimes of DS transport for various linear gain (a) $\delta=0.04$, (b) $\delta=0.05$, and (c) $\delta=0.06$, when spectral filtering $\beta=0.01$ and nonlinear gain $\epsilon=0.8\epsilon_p$. Results are obtained from the numerical solution of CGLE with $V_s=V_l=5$, $\nu=0.1$, $\beta=0.01$, $N_T=41$ and $dt=0.0025$. We use the initial condition in the form of Eq.~(3) in main text. Panels (d)-(f): Peak intensity $|A|^2_\textrm{max}(z)$ and participation ratio $r(z)$ of the instantaneous soliton state in (a)-(c), respectively.}\label{Fig:Suppfig4}
\end{figure}

\section{Non-Hermitian linear Bloch Bands and Chern number}\label{AppendixD}

Here we present (i) calculations of the band structures associated with the linear non-Hermitian Hamiltonian of CGLE and their Chern numbers, (ii) projections of the truncated soliton state onto the Bloch states.

As described in the main text, the linear part associated with CGLE (1) in the main text is 
\begin{equation}
i\frac{\partial A}{\partial z}=H_0A\label{SingleSchro}
\end{equation}
with the linear non-Hermitian Hamiltonian 
\begin{align}\label{Eq:NonHermitianH}
	H_0=\left(i\beta-\frac{1}{2}\right)\frac{\partial^2}{\partial t^2}+i\delta+V_{\text{ext}}(t,z).
\end{align}

We now describe how to obtain the Bloch bands associated with $H_0$ in the space-time domain. First, since $V_{\text{ext}}(t,z)$ is periodic in time with the periodicity $T=1$, according to the Bloch's theorem $A(t,z)=e^{-i\omega t}u(t,z)$ with the quasi-frequency $\omega\in[0,2\pi/T)$ and $u(t+T,z)=u(t,z)$, we expand 
\begin{equation}
A(t,z)=e^{-i\omega t}\sum_{n=-\infty}^{+\infty}c_n(z)e^{inQt},
\end{equation}
 where $Q=2\pi/T$ and $c_n(z)$ is the expansion coefficient. Inserting this expansion into Eq.~(\ref{SingleSchro}) and denoting $\mathbf{c}=(\cdots,c_{-1},c_0,c_1,\cdots)^T$, we formally obtain 
\begin{align}
	i\frac{\partial}{\partial z}\mathbf{c}=H_0(\omega,z)\mathbf{c}.
\end{align}
Here, the effective Hamiltonian matrix $H_0(\omega,z)$, with parametric dependence on the propagation distance $z$, satisfies $H(\omega,z+L)=H(\omega,z)$ with the spatial periodicity $L=\pi/\nu$. In the numerical calculation, the expansion is usually truncated at $\pm n_\textrm{max}=\pm M$. This results in the truncated matrix $H(\omega,z)$ with the dimension $(2M+1)\times (2M+1)$, which is given by
\begin{widetext}
	\begin{equation}
		H(\omega,z)=\frac{1}{4}\left(\begin{array}{cccccccccc}
			\ddots&&&&\\
			\cdots&h_{-2}&-{V_l}\xi(z)&-{V_s}&&\\
			\cdots&-{V_l}\xi^\star(z)&h_{-1}&-{V_l}\xi(z)&-{V_s}\\
			&-{V_s}&-{V_l}\xi^\star(z)&h_0&-{V_l}\xi(z)&-{V_s}\\
			&&-{V_s}&-{V_l}\xi^\star(z)&h_1&-{V_l}\xi(z)&\cdots\\
			&&&-{V_s}&-{V_l}\xi^\star(z)&h_2&\cdots\\
			&&&&&&\ddots&
		\end{array}\right)_{(2M+1)\times(2M+1)}.
	\end{equation}
\end{widetext}
Here, $h_n=(2-4i\beta)(\omega-nQ)^2+4i\delta-2V_s-2V_l$ for $n=-M,\cdots,M$, and $\xi(z)=\exp(2i\nu z)$. In practice, $M=10$ is sufficient for calculating the Chern numbers of the band structures. 

Second, we diagonalize $H_0(\omega,z)$ to find the adiabatic right eigenstate $|R_\alpha(\omega,z)\rangle$ and corresponding eigenvalue $\epsilon_\alpha(\omega,z)$ ($\alpha=1,...2M+1$). Since $H_0$ is non-Hermitian, the eigenvalues are generically complex. These eigenvalues form band structures in the two-dimensional parameter space $(\omega,z)$. The Chern number $C_\alpha$ associated with the $\alpha$-band is then defined as~\cite{Huitao2018,Yao2018}
\begin{eqnarray}
	C_\alpha=\frac{i}{2\pi}\int_0^Ldz\int_{0}^{2\pi/T}d\omega\Big[\langle\partial_zR_\alpha|\partial_\omega R_\alpha\rangle-\langle\partial_\omega R_\alpha|\partial_zR_\alpha\rangle\Big].\label{C}
\end{eqnarray}

Following the above procedures, one reproduces the known results in the case of linear topological pump (e.g. Refs.~\cite{Wang2013,Jurgensen2021}) when $\delta=\beta=\epsilon=0$. For the system parameters in Figs.~2 and 3 of the main text, the Chern number for the lowest three bands are all obtained as $C=\{1,-1,1\}$. In Fig.~\ref{FigSupp3}, we show the complex band structures of the lowest two Bloch bands (ordered according to the real components of the eigenvalues). 

\begin{figure}[t]
	\centering
	\includegraphics[width=0.8\columnwidth]{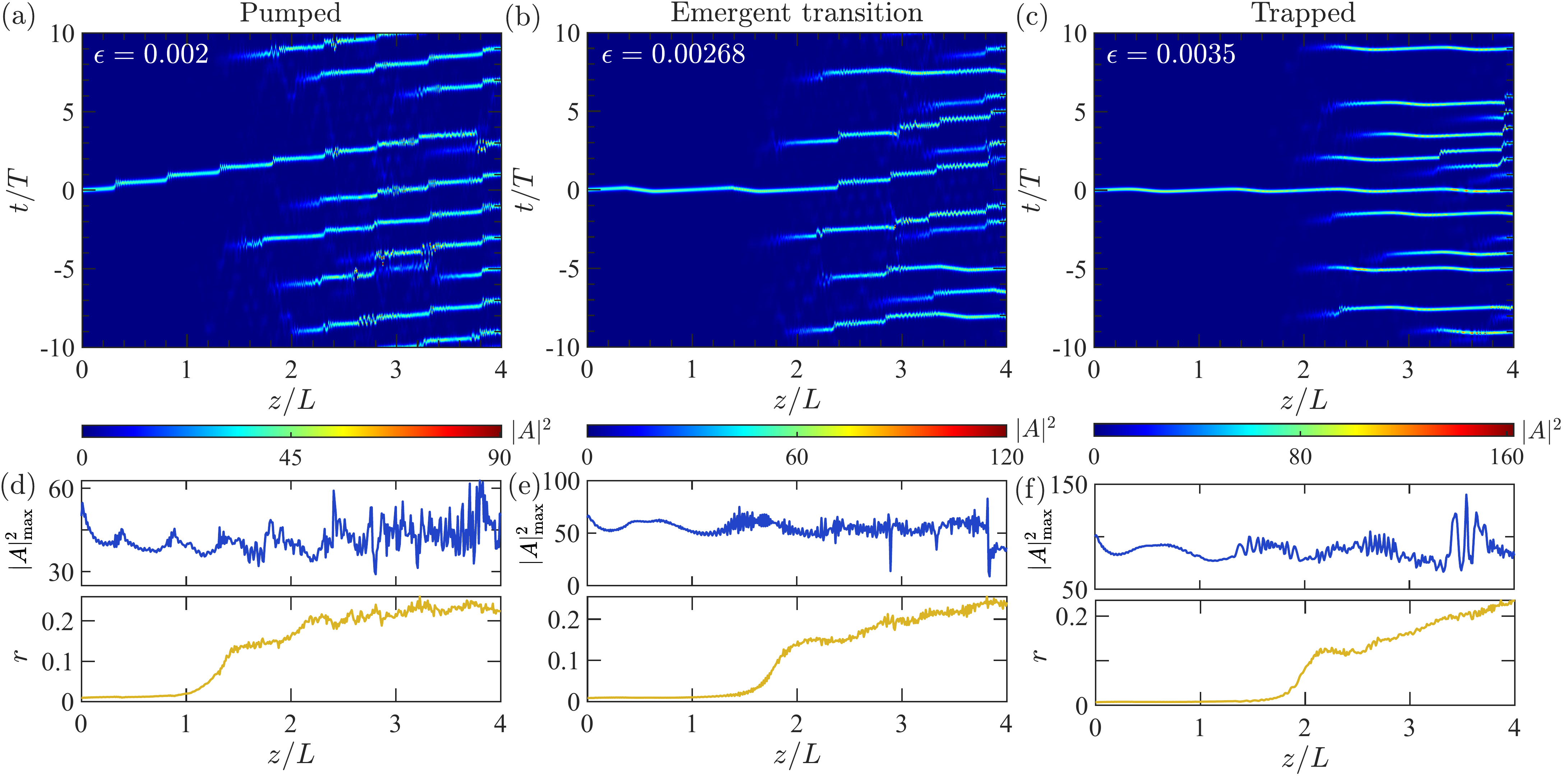}
	\caption{Pumped, emergent transition, and trapped regimes of DS transport for various nonlinear gain (a) $\epsilon=0.002$, (b) $\epsilon=0.00268$, (c) $\epsilon=0.0035$, when $\beta=0.01$ and $\delta=0.1$. Numerical results are obtained for CGLE with $V_s=V_l=5$, $N_T=41$, $dt=0.0025$, and the initial condition Eq.~(3) in the main text. }\label{Fig:Suppfig5}
\end{figure}

Finally, we describe how to obtain the projection of the truncated soliton in the time domain. By discretizing the time domain and writing $\frac{\partial^2}{\partial t^2}A(t,z)\equiv \frac{A(t+dt,z)-2A(t,z)+A(t-dt,z)}{dt^2}$, we recast $H_0(t,z)$ in Eq.~(\ref{Eq:NonHermitianH}) into a discrete form. Then, we diagonalize this discretized Hamiltonian to obtain the form of the eigenstate $|n\rangle$ of the $n_\text{th}$ level in time domain. The projection of the renormalized truncated soliton state $A_p$ onto the $n_{th}$ level is $
	p_n(z)=\frac{|\langle n|A_p\rangle|^2}{\langle A_p|A_p\rangle}$. For line-gapped non-Hermitian bands, the instantaneous projection of the truncated soliton state on the $\alpha_{th}$ band is given by 
\begin{equation}
	P_\alpha=\sum_{n=(\alpha-1) N_T+1}^{\alpha N_T}p_n.
\end{equation}

\begin{figure}[t]
	\centering
	\includegraphics[width=0.3\columnwidth]{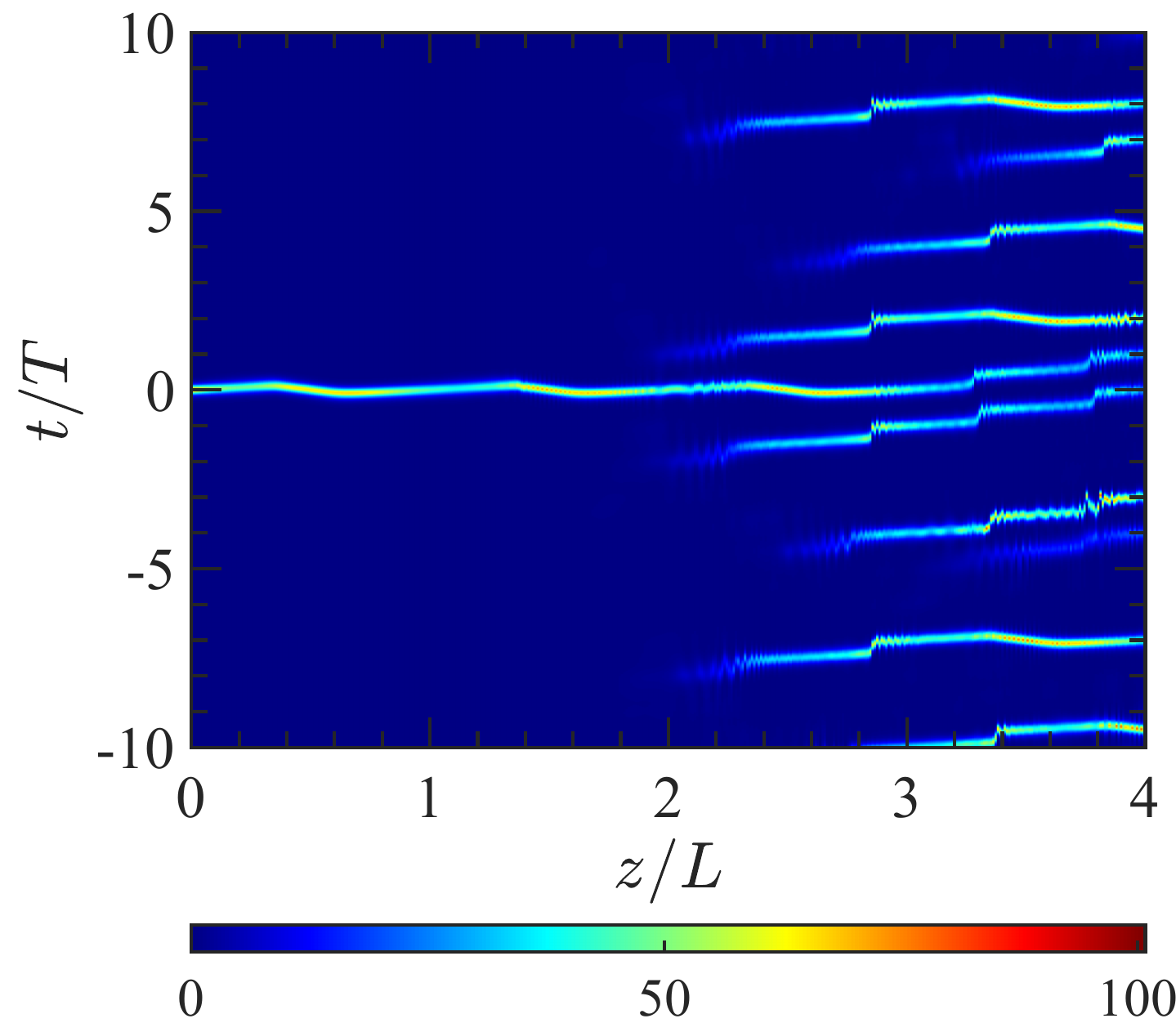}
	\caption{Numerical solution of CGLE with $dt=0.00125$ and $V_s=V_l=5$, $\delta=0.1$, $\beta=0.02$, $\epsilon=0.8\epsilon_p$, $N_T=41$, $dt=0.00125$, and with the initial condition Eq.~(3) in main text.}\label{Fig:Suppfig6}
\end{figure}
\begin{figure}[t]
	\centering
	\includegraphics[width=0.85\columnwidth]{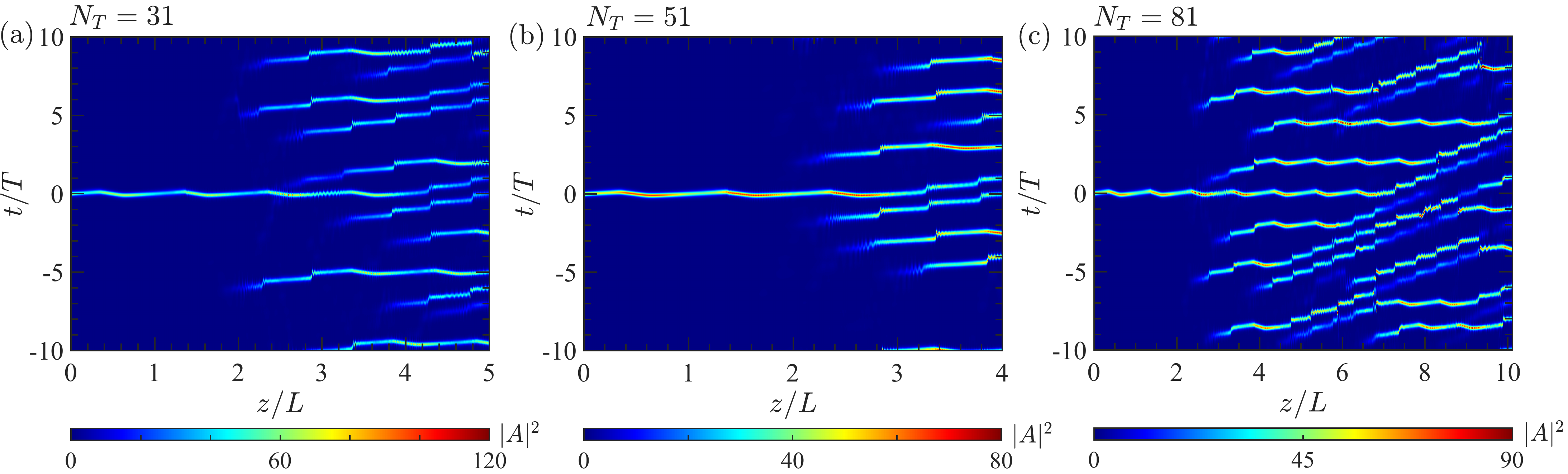}
	\caption{Dynamically emergent phase transition under different system size (a) $N_T=31$, (b) $N_T=51$, and (c) $N_T=81$. The system parameters are $\delta=0.1$, $\beta=0.02$, and $\epsilon=0.8\epsilon_p$.}\label{Fig:Suppfig7}
\end{figure}

\section{Generality of results}\label{AppendixE}
In the main text, we have shown the four regimes of DS behavior when changing spectral filtering $\beta$. In this section, we show that our key results, including the quantized transport of DS and the dynamically emergent phase transitions, are generic by considering various choices of $\delta$, $\epsilon$,  $dt$, and system size $N_T$. For this purpose, the collapsed regime will not be shown in the plots. 

In Fig.~\ref{Fig:Suppfig4}, we show the DS behavior by varying linear gain $\delta$, while we fix spectral filtering parameter $\beta=0.01$ and choose $\epsilon=0.8\epsilon_p$. When $\delta$ increases, Fig.~\ref{Fig:Suppfig4} reveals the pumped regime, the dynamically emergent phase transition regime, and the trapped regime. The corresponding peak intensity and participation ratio $r$ in various regimes are also shown. These results are in accordance with our analysis in the main text. Similar three regimes also show up in Fig.~\ref{Fig:Suppfig5}, where nonlinear gain $\epsilon$ is increased, while $\beta$ and $\delta$ are fixed.

In our numerical simulation, we used $dt=0.0025$. Choosing smaller $dt$ does not lead to notable quantitative changes, as shown in Fig.~\ref{Fig:Suppfig6} for the numerical solution of CGLE with $dt=0.00125$ while other parameters are the same as Fig.~2(b) of the main text. The dynamically emergent phase transition is still manifest.

In Fig.~\ref{Fig:Suppfig7}, we show the numerical solutions of CGLE for various size $N_T$ of the temporal domain. In all cases, dynamically emergent phase transitions are observed in the behavior of the primary soliton, although the critical point of this transition may vary.